\documentclass[conference]{IEEEtran}
\IEEEoverridecommandlockouts
\usepackage{cite}
\usepackage{amsmath,amssymb,amsfonts}
\usepackage{algorithmic}
\usepackage{graphicx}
\usepackage{textcomp}
\usepackage{romannum}
\usepackage{url}
\usepackage{booktabs}
\usepackage{threeparttable}
\usepackage[caption=false]{subfig}
\usepackage{bm}
\usepackage{tabularx}
\usepackage{multirow}
\usepackage[table,xcdraw]{xcolor}

\usepackage{booktabs}
\usepackage{cleveref}

\def\BibTeX{{\rm B\kern-.05em{\sc i\kern-.025em b}\kern-.08em
    T\kern-.1667em\lower.7ex\hbox{E}\kern-.125emX}}

\begin{document}

\title{Self-Supervised Pretext Tasks for Alzheimer's Disease Classification using 3D Convolutional Neural Networks on Large-Scale Synthetic Neuroimaging Dataset}
% 1\textsuperscript{st} 2\textsuperscript{nd} 3\textsuperscript{rd}
\author{
\IEEEauthorblockN{Chen ZHENG\IEEEauthorrefmark{1}}
\IEEEauthorblockA{\textit{Computing and Mathematical Sciences} \\
\textit{University of Waikato}\\
Hamilton, New Zealand \\
zhengchenloot@gmail.com}
}

% \and
% \IEEEauthorblockN{Bernhard Pfahringer}
% \IEEEauthorblockA{\textit{Computing and Mathematical Sciences} \\
% \textit{University of Waikato}\\
% Hamilton, New Zealand \\
% bernhard.pfahringer@gmail.com}

\maketitle

\begingroup\renewcommand\thefootnote{\textsection}
\footnotetext{Chief Supervisor Prof. Bernhard Pfahringer, Co-Supervisor Dr. Michael Mayo, Computing and Mathematical Sciences, University of Waikato, Hamilton, New Zealand}

\begin{abstract}
Structural magnetic resonance imaging (MRI) studies have shown that Alzheimer’s Disease (AD) induces both localised and widespread neural degenerative changes throughout the brain. However, the absence of segmentation that highlights brain degenerative changes presents unique challenges for training CNN-based classifiers in a supervised fashion. In this work, we evaluated several unsupervised methods to train a feature extractor for downstream AD vs. CN classification. Using the 3D T1-weighted MRI data of cognitive normal (CN) subjects from the synthetic neuroimaging LDM100K dataset, lightweight 3D CNN-based models are trained for brain age prediction, brain image rotation classification, brain image reconstruction and a multi-head task combining all three tasks into one. Feature extractors trained on the LDM100K synthetic dataset achieved similar performance compared to the same model using real-world data. This supports the feasibility of utilising large-scale synthetic data for pretext task training. All the training and testing splits are performed on the subject-level to prevent data leakage issues. Alongside the simple preprocessing steps, the random cropping data augmentation technique shows consistent improvement across all experiments. \par
\end{abstract}

% \item The subject-level train-test-split fundamentally eliminates the risk of data leakage.
%\item Our 3D CNN architecture can fully utilise the spatial information of 3D MRI scans, which completely eliminates the need of hyperparameter tuning for the number of slices and the number or the location of patches.
%\item Our approach explicitly incorporate the subject's meta information age into the model learning process, which brings more prior knowledge into the feature extractor.
%\item Our approach can achieve comparable performance in the absence of complicated preprocessing techniques such as scan registration and scan skull-stripping, therefore greatly reducing the overall complexity.
%\item Unlike other studies that rely heavily on data augmentation techniques for performance, the classification results demonstrated the robustness of our approach.


\section{Introduction}
Automated medical image analysis systems have been built since it has been possible to scan and load medical images into a computer. From the 1970s to the 1990s, sequential application of low-level pixel processing (edge and line detector filters, region growing) and mathematical modelling (fitting lines, circles and ellipses) were used for medical image analysis \cite{litjens2017survey}. In the same period, artificial intelligence concepts were implemented with many if-then-else statements to form rule-based image processing systems \cite{raya1990low}. \par

Although there is a shift from systems that are completely designed by humans to systems that are trained by computers using example data from which features are extracted, the extraction of discriminant features from the images is still done by human researchers \cite{litjens2017survey}. Conventionally, constructing such machine learning systems requires meticulous engineering and considerable domain knowledge to transform the raw data (i.e. the pixel values of an image) into discriminative features, such as the shape, colour, and/or texture roughness as well as their combinations. The learning system then determines the optimal decision boundary in the feature space so that it can detect patterns (e.g. tumor tissue) in the raw data. However, this transformation relies on handcrafted features, thus the conventional machine-learning techniques are limited in their ability to process raw data. Logically, one possible next step is to let the computer learn such features for discrimination directly from raw data. \par

Recently, the Machine Learning (ML) field has received enormous attention due to the exciting breakthrough of the Deep Learning technique\cite{lundervold2019overview}. Deep learning is a sub-field of machine learning, but it is different from traditional machine learning in that features are learned from the raw data. Deep learning aims to discover high-level data representations by utilising hierarchical architectures \cite{lecun2015deep}. For example, a cat's different combinations of shape, colour, and texture can be represented at a higher abstract level. Deep learning computational models consist of multiple processing layers to learn such representations of data with multiple levels of abstraction. With the composition of enough such layers, appropriate discriminative features can be learned as high-level representations in a latent space. The reason is that higher levels of representation amplify aspects of the input that are important for discrimination and suppress irrelevant variations \cite{lecun2015deep}. \par

The majority of modern deep learning architectures are based on feedforward neural networks. They are built upon many layers of non-linear processing units for feature extraction. Data is fed into the input layer and propagates through the network to reach a final output. An error value is obtained by comparing the network output against the expected value. The training procedure propagates the error backwards and updates the weights of the network to minimise the error \cite{lecun2012efficient}. This supervised learning/training process is repeated a given number of epochs to optimise the weights. \par

Supervised learning is the most common training technique when the data are well-labelled. The idea is to learn the relevance of different features by fitting the known outcome. CNN-based classifiers in the literature show promising performance in some tasks. However, achieving such performance requires a vast amount of well-labelled data. The collection of labelled data can be costly and extremely difficult or even impossible to obtain. Therefore, self-supervised learning techniques can be very beneficial. \par

Self-supervised learning belongs to the unsupervised learning method family which can be applied to suitable data formats (e.g. images) where labels are not available. Generally, computer vision pipelines that employ self-supervised learning involve performing two tasks, a pretext task and a target task. The pretext task is the self-supervised learning task solved to learn meaningful representations, with the aim of using the learned representations or model weights obtained in the process, for the target task. This is based on the assumption that the unlabelled data can be represented in a semantic and structurally meaningful way by learning the pretext task. The pretext task is generally designed by fabricating artificial labels from unlabelled data without human annotation. The target task can be anything like a classification or detection task that requires a vast amount of well-labelled data. \par

Age is one of the primary risk factors for neurodegenerative diseases, especially dementia and its most common form, Alzheimer’s Disease (AD) \cite{winblad2016defeating}. AD causes an irreversible progressive impairment of memory and cognitive functions that considerably disrupts a patient's daily life. Because of the aging population, the cost of care for AD is expected to increase from the current 47 million to 152 million by 2050 \cite{alzheimer20192019} just in the US. In view of the lack of progress in developing effective treatment for AD \cite{winblad2016defeating} and the rapidly increasing costs of medical care and associated socioeconomic impact, defeating AD is a priority for science and society. \par

Due to human brain inaccessibility, structural magnetic resonance imaging (MRI) is a widely used neuroimaging technique to assess the brain of an individual. Fig.~\ref{fig:3d_planes} is an exmaple of the neuroimaging  planes of a 3D MRI scan. In the context of neurodegenerative diseases, particularly AD, brain changes are consistently found in the form of atrophy (e.g. tissue loss or shrinkage) compared to a population of healthy brains, as shown in Fig.~\ref{fig:brain_atrophy}. Such AD disease progression is not limited to specific brain regions or sites (e.g. brain tumor) \cite{bateman2012clinical}. Well-segmented data is often unavailable in the medical domain, particularly in the field of AD neuroimaging \cite{adni, oasis3}. Therefore, AD neuroimaging data are often only categories into possible diagnostic groups in clinical practice. \par

The task of differentiating subjects with AD from cognitively normal (CN) subjects is popular in the literature. As shown in Fig.~\ref{fig:cn_mci_ad}, before the development of AD, subjects go through an early stage called mild cognitive impairment (MCI) during which their cognitive functions are deteriorated. Subjects with MCI may remain cognitive stable or subsequently progress to AD. Therefore, identifying MCI subjects from CN subjects is another task of interest. \par

\begin{figure}[htbp]
\centerline{\includegraphics[width=0.45\textwidth]{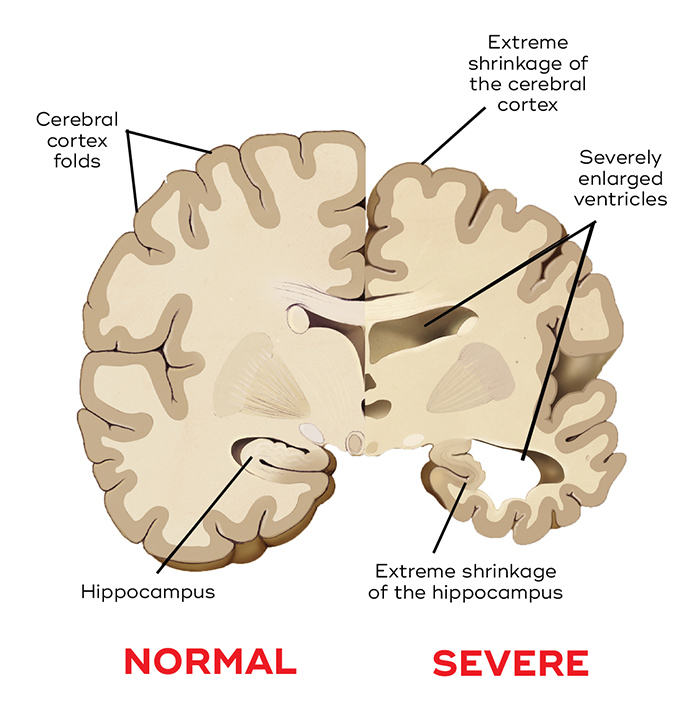}}
\vspace{-0.5cm}
\caption{Comparison of brain tissue between a normal brain (left) and severe brain atrophy caused by Alzheimer's Disease (right) from a coronal (frontal) point of view. This figure is copied from website \cite{brainshrinkage2018}.}
\label{fig:brain_atrophy}
\end{figure}

\begin{figure}[htbp]
\centerline{\includegraphics[width=0.45\textwidth]{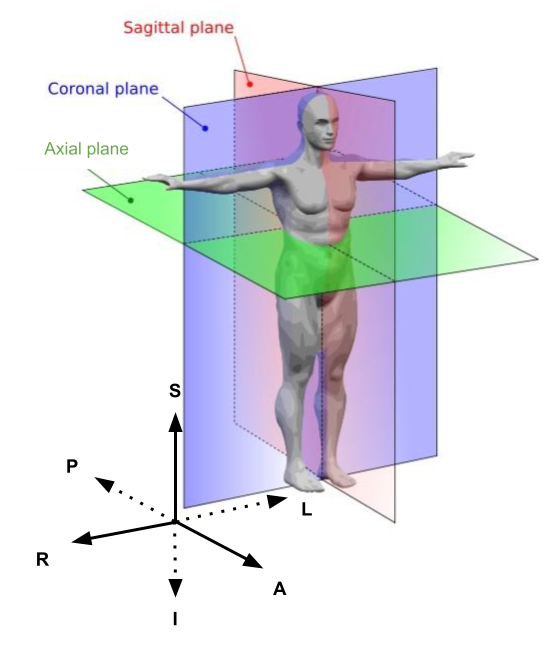}}
\vspace{-0.5cm}
\caption{MRI is defined by the plane (direction) of the image that is taken. Typically, three planes are used to describe the standard anatomical position of a human \cite{mri_planes}. The basic orientation terms for a MRI of the body: from the inferior (I) to superior (S) is the \emph{axial} plane; from the left (L) to right (R) is the \emph{sagittal} plane; and from the anterior (A) to posterior (P) is the \emph{coronal} plane. Figure source \cite{planes}.}
\label{fig:3d_planes}
\end{figure}

\begin{figure}[htbp]
\centerline{\includegraphics[width=0.5\textwidth]{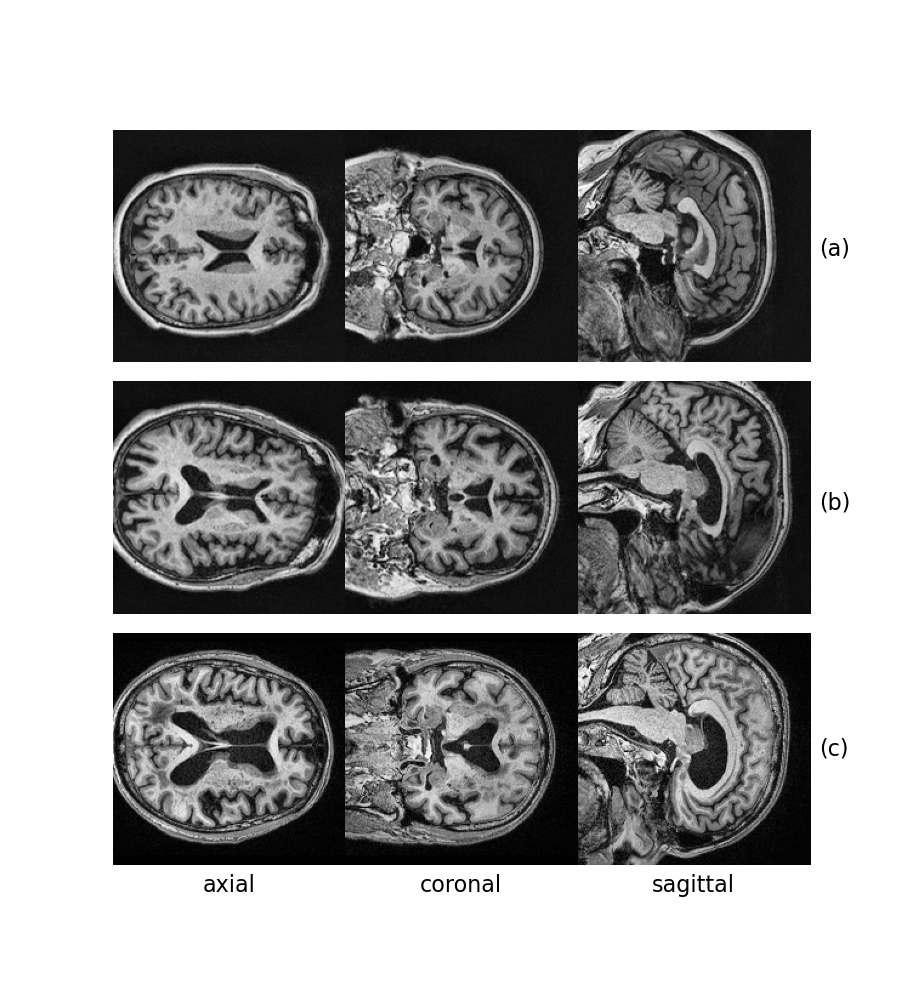}}
\vspace{-1cm}
\caption{Three examples of T1-weighted MRI scans slices of subjects from the OASIS-3 \cite{oasis3} dataset in three planes: axial, coronal and sagittal. Row (a) shows images from a healthy subject whereas row (b) comes from a MCI subject with obvious brain changes in all planes. Lastly, row (c) reveals the severe brain tissue loss of an Alzheimer's subject. }
\label{fig:cn_mci_ad}
\end{figure}

%  The idea is that if the predicted age of subject is higher than the actual age, this implies the brain is , results in a "older-appearing" MRI.
% Medical studies \cite{cutler2004involvement,klunk2004imaging} found that AD subjects have a significantly “older-appearing” brain compared with normal aging subjects.

\section{Related Work}
As shown in Fig.\ref{fig:brain_atrophy} and Fig.\ref{fig:cn_mci_ad}, the brain tissue loss is most obvious between the CN and AD classes, therefore, the class difference is most distinguishable on structural MRI scans. As a result, the classification between Alzheimer's Disease (AD) and cognitively normal (CN) subjects is the most studied task in the literature. Before the development of dementia, patients go through a phase called mild cognitive impairment (MCI) in which they experience the deterioration of certain cognitive functions, such as having difficulty reasoning. MCI patients can be subdivided into stable, progressive and converting stages or early and old stages, denoted as sMCI, pMCI or cMCI, EMCI and LMCI respectively. Identifying this cognitive stage of a patient is another task of interest. \par

Inspired by the success of natural image classification using 2D CNNs (e.g. ResNet \cite{resnet} and VGGNet \cite{vggnet}), extracting 2D slices from 3D MRI volumes is a popular choice of input in many studies. Many studies report state-of-the-art classification accuracy on various datasets. However, many of the proposed approaches are suspected to have data leakage or biased evaluation metrics. It is speculated that the authors are primarily from medical domains and are not experienced in machine learning \cite{review}. The rapid adoption of machine learning methods in a certain domain could lead to errors. Moreover, the classification results are often unclear about slice-level or subject-level accuracy. \par

In the literature, self-supervised learning methods using image rotation classification have succeeded in learning useful representations for downstream tasks. The attraction of Rotation is simplicity and effectiveness. The most widely used rotation-based task is to identify the rotation degree of input images (e.g., 0◦, 90◦, 180◦, and 270◦). By solving Rotation, the models can capture information about object shapes therefore achieving competitive performance in the downstream tasks \cite{gidaris2018unsupervised, zhai2019s4l, yamaguchi2021image, oh2019classification}. Study \cite{yamaguchi2021image} also incorporated the image enhancement (e.g. brightness and contrast) as prediction targets in a multi-head training configuration. \par

AutoEncoder-based representation learning is another popular branch of self-supervised methods. Many studies have achieved promising results. Study \cite{chen2024context} proposed an encoder-regressor-decoder architecture to train feature extractors. The encoder and regressor generate the representation of a masked image while the decoder reconstructs the masked patches. Study \cite{huang2021self} introduced a transformation-augmented AutoEncoder architecture that reconstructs both the original input image and the applied transformations for anomaly detection. Study \cite{he2021autoencoder} proposed an AutoEncoder-based transfer learning scheme to reduce domain shift. The authors reported significant performance improvement upon validation using T1 and T2 MRI images. \par

AutoEncoders are also popular pretext tasks for AD classification. Study \cite{li2017alzheimer} proposed a classification approach consisting of two feature extractors: part one is a deep 3D CNN, whereas part two has multiple 3D CNN AutoEncoders (AEs).  The AutoEncoders are pre-trained for sMRI reconstruction followed by a fine-tuning for AD vs. CN classification. Study \cite{venugopalan2021multimodal} used 3D CNN for imaging data and stacked Denoising AutoEncoders for clinical and genetic data to extract features for fusion followed by classification. \cite{wu20223d} proposed a transfer learning and self-supervised method to classify AD on the T1w sMRI modality using 2D CNN AutoEncoder. Study \cite{zhao2021longitudinal} proposed a self-supervised learning approach for AD classification using AutoEncoders. The autoencoder is pre-trained on a synthetic dataset with image pairs to learn the development of the AD over time. \par

A very recent publication \cite{kapoor2023leakage} highlights this issue which is apparently present in all application areas of machine learning. The common data leakage found in the field of Alzheimer’s Disease classification are:
\begin{itemize}
    \item Leakage A: Wrong Data Split. The data split is not performed on the subject-level when defining the train, test and validation subset, resulting in data from the same subject appearing in multiple subsets. This problem can occur when 2D patches or 2D slices are extracted from a 3D image, or when 3D images of the same subject are available at multiple time points.
    \item Leakage B: Late split. Procedures such as data augmentation, feature selection or pre-training tasks are performed before subject-level training, validation and testing splits. For example, the generated image of the same subject could appear in several subsets if the augmentation is performed before the subject-level split.
    \item Leakage C: Biased transfer learning. Transfer learning can cause data leakage if the source and target domains overlap. For example, the CN subjects used in a pre-training task of an AutoEncoder are also used in the downstream task of CN vs. MCI or CN vs. AD classification.
\end{itemize}

For example, \cite{li2017alzheimer} only used the baseline visit per subject, there is no data leakage detected. Similarly, \cite{venugopalan2021multimodal} is only using one MRI image per subject, so there is no data leakage. The authors of \cite{wu20223d}  are aware of the data leakage problem and clearly performed subject-level train-test-split. However, in study \cite{zhao2021longitudinal, orouskhani2022alzheimer, savacs2022detecting, goenka2022alzvnet, kang2021multi}, the lack of clarification of train-test-split raises the concern of type B data leakage in their evaluation. \par

%By leveraging image enhancement that modifies the appearances of images, the model can learn the information of textures. The authors reported improved downstream classification performance across various datasets and tasks.

\section{Method}
\subsection{Subjects}
This work utilised three publicly available brain neuroimaging datasets for Alzheimer’s disease (AD), including two real-world datasets: OASIS and ADNI and a synthetic dataset LDM100K. Both the OASIS and ADNI focus on Alzheimer’s disease, but none of them has detailed segmentation for the disease lesion, which would indicate the regions in an organ or tissue that have suffered damage caused by the disease. A brief summary of the datasets is shown in Table.\ref{tab:datasets}. \par

Please note that: *The Mini-Mental State Examination (MMSE) measures general cognitive status, with scores ranging from 0 (severe impairment to 30 (no impairment). **The Clinical Dementia Rating (CDR) is a numeric scale used to quantify the severity (stages) of dementia in clinical practice. Clinical Dementia Rating Assignment Qualitative equivalences are as follows: 0 is none; 0.5 is very mild; 1 is mild; 2 is moderate and 3 is severe. \par

\begin{table}[htbp]
\resizebox{\columnwidth}{!}{%
\begin{tabular}{|l|l|l|l|l|l|l|l|}
\hline
\textbf{Dataset} & \textbf{Classes} & \textbf{\#Subjects} & \textbf{\#Scans} & \textbf{$MMSE^{*}$} & \textbf{$CDR^{**}$} & \textbf{Sex} & \textbf{Age} \\ \hline
ADNI & \begin{tabular}[c]{@{}l@{}}AD\\ MCI\\ CN\end{tabular} & \begin{tabular}[c]{@{}l@{}}192\\ 398\\ 229\end{tabular} & \begin{tabular}[c]{@{}l@{}}530\\ 1126\\ 877\end{tabular} & \begin{tabular}[c]{@{}l@{}}23.3 +/- 2.1\\ 27.0 +/- 1.8\\ 29.1 +/- 1.0\end{tabular} & \begin{tabular}[c]{@{}l@{}}2\\ 1, 0.5\\ 0\end{tabular} & \begin{tabular}[c]{@{}l@{}}Male 1,426\\ Female 1,316\end{tabular} & 55 - 94 \\ \hline
OASIS-3 & \begin{tabular}[c]{@{}l@{}}AD\\ MCI\\ CN\end{tabular} & \begin{tabular}[c]{@{}l@{}}104\\ 389\\ 605\end{tabular} & \begin{tabular}[c]{@{}l@{}}237\\ 1131\\ 2021\end{tabular} & \begin{tabular}[c]{@{}l@{}}15.0 +/- 5.8\\ 24.9 +/- 3.3\\ 29.1 +/- 1.2\end{tabular} & \begin{tabular}[c]{@{}l@{}}2\\ 1, 0.5\\ 0\end{tabular} & \begin{tabular}[c]{@{}l@{}}Male 487\\ Female 611\end{tabular} & 42 - 95 \\ \hline

LDM-100k & CN & 100,000 & 100,000 & N/A & N/A & N/A & 44 - 82 \\ \hline
\end{tabular}%
}
\caption{A summary of the neuroimaging datasets.}
\label{tab:datasets}
\end{table}

\subsection{Subject Selection}
In this work utilise synthetic data from the LDM-100k dataset to train the AutoEncoder. For testing, we selected AD and CN subjects from the OASIS-3 and the ADNI dataset. The overview of the experimental settings for this chapter approach is depicted in Fig.\ref{fig:Age_Split_1}. The ``\#" shows the number of subjects and scans, e.g. there are 605 CN Subjects with 1,978 scans from the OASIS-3 dataset. \par

\begin{figure}[htbp]
\centering
\includegraphics[width=0.975\linewidth]{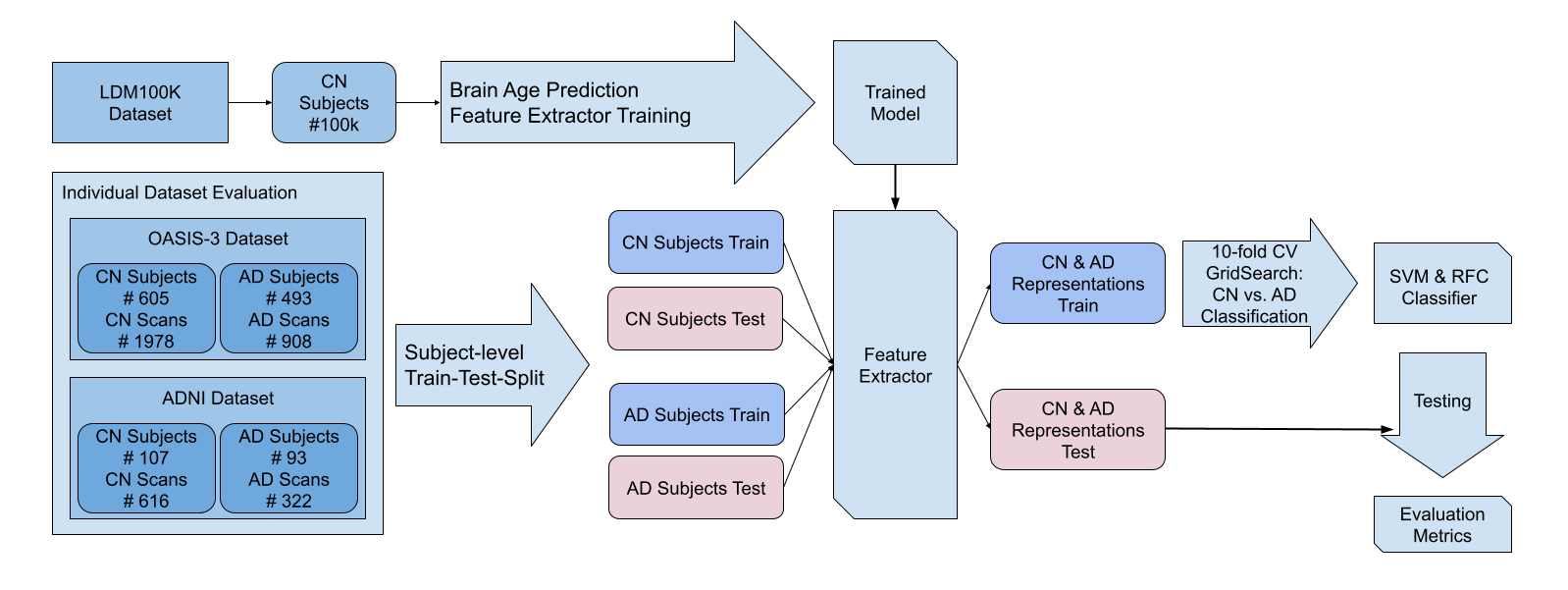}
\hfill
\caption{Overview of the train-test-split.} 
\label{fig:Age_Split_1}
\end{figure}

\subsubsection{Training Data}
As shown in Fig.\ref{fig:Age_Split_1}, the training data is selected from the LMD100K dataset. It is a large dataset recently created by \cite{ldm100k}. The authors utilised the latent diffusion models to generate synthetic images from high-resolution 3D brain images. The models are trained to learn about the probabilistic distribution of brain images while conditioned on covariables, such as age, sex, and brain structure volumes. \par

Given the quantity of data points in LDM100k, it can be used to train large and complex models. However, the high-resolution 3D images are computationally expensive. The previous chapter used a 10-fold cross-validation setup to obtain multiple models. Due to hardware limitations, the training process was reduced to 1-fold for this data. \par

\subsubsection{Testing Data}
As shown in Fig.\ref{fig:Age_Split_1}, the testing data are chosen from the OASIS-3 dataset and ADNI dataset. The subject IDs are inconsistent in these two datasets, and there is no documentation regarding the generation of the subject IDs. Therefore, evaluation is performed independently on both datasets to eliminate the change of data leakage. The number of subjects used for testing is significantly increased compared to the previous chapter due to the involvement of the LDM100k dataset for training. \par

For the OASIS dataset, the CDR-based subject selection is refined for this chapter and the following ones. More specifically, each sMRI scan has a timestamp, as well as each CDR assessment. The problem is that they are performed on different dates. We associated each sMRI scan with a CDR score that is closest to the date of the scan. A subject is selected based on the highest CDR score. Then the lower-scored sMRI scans are discarded for a better representation of the real-world situation. For example, a subject is added to the AD group if their CDR score is 2 or 3. The subjects with CDR 0.5 and 1 are also included in the AD group to counter the data imbalance issue. Therefore, the CN group only consist of subjects with CDR 0. In total, there are 605 CN subjects and 493 AD subjects, yielding 1978 and 908 sMRI images, respectively.  \par

From the ADNI database, CN and AD subjects are selected based on the labels given by the ADNI database host. For comparison purposes, the sMRI data are chosen from their screening visit. This is a popular selection criterion in the literature as the screening visit includes more amount of sMRI data per subject than any of the follow-up visits. In total, there are 107 CN subjects and 93 AD subjects, yielding 616 and 322 sMRI images, respectively. \par

A subject-level train-test-split is carried out during the evaluation for both OASIS-3 and ADNI datasets to address the data leakage issue. However, as there might be more than one sMRI scan per subject, a data leakage type A might occur during the 10-fold cross-validation. For example, the sMRI images of one subject could be split into both the training set and the validation set. To avoid that, a subject ID grouped split is configured for the cross-validation process. \par

\subsection{Preprocessing}
The whole brain scan size of $256\times256\times256$ voxels is not GPU memory efficient due to the large amount of black voxels on the scan boarder. Therefore, resizing all 3D MRI scans to $192\times192\times192$ voxels is a logical choice since it still maintains all brain tissue voxels without consuming extensive GPU memory. Preliminary results show that smaller scan sizes (e.g. $156\times156\times156$ and $128\times128\times128$) significantly accelerate the training, however, they cannot yield good results. The possible reason could be that the subtle yet informative structural brain changes caused by AD are lost during drastic downsampling and the resizing opration might produce artifacts in the image. \par

As previously shown \cite{review}, preprocessing has significant impact on the performance of brain age estimation and AD classification. Therefore, for establishing a baseline, this study only employed minimum steps of preprocessing. Sequentially, each MRI scan undertakes a pipeline (shown in Fig.~\ref{fig:preprocessing}) of:
\begin{enumerate}
  \item Resizing: To utilise the 3D scans that have different shapes, also to remove the excessive background image volume outside the brain tissue, all of them are resampled to a pre-selected shape $192\times192\times192$.

  \item Max-Min Intensity Normalisation: To eliminate the inconsistency of 3D scan voxel intensities, e.g. under or over exposure, the range of intensities is normalised by $I_{norm}=\frac{I-I_{min}}{I_{max}-I_{min}}$.
  
  \vspace{2px}
  \item Contrast Limited Adaptive Histogram Equalisation \cite{clahe}: To enhance the contrast between different brain tissue types, adaptive histogram equalisation is performed on each 3D scan. The contrast limited variation is chosen to reduce the overamplification of noise.
\end{enumerate}

\begin{figure}[htbp]
\centerline{\includegraphics[width=0.5\textwidth]{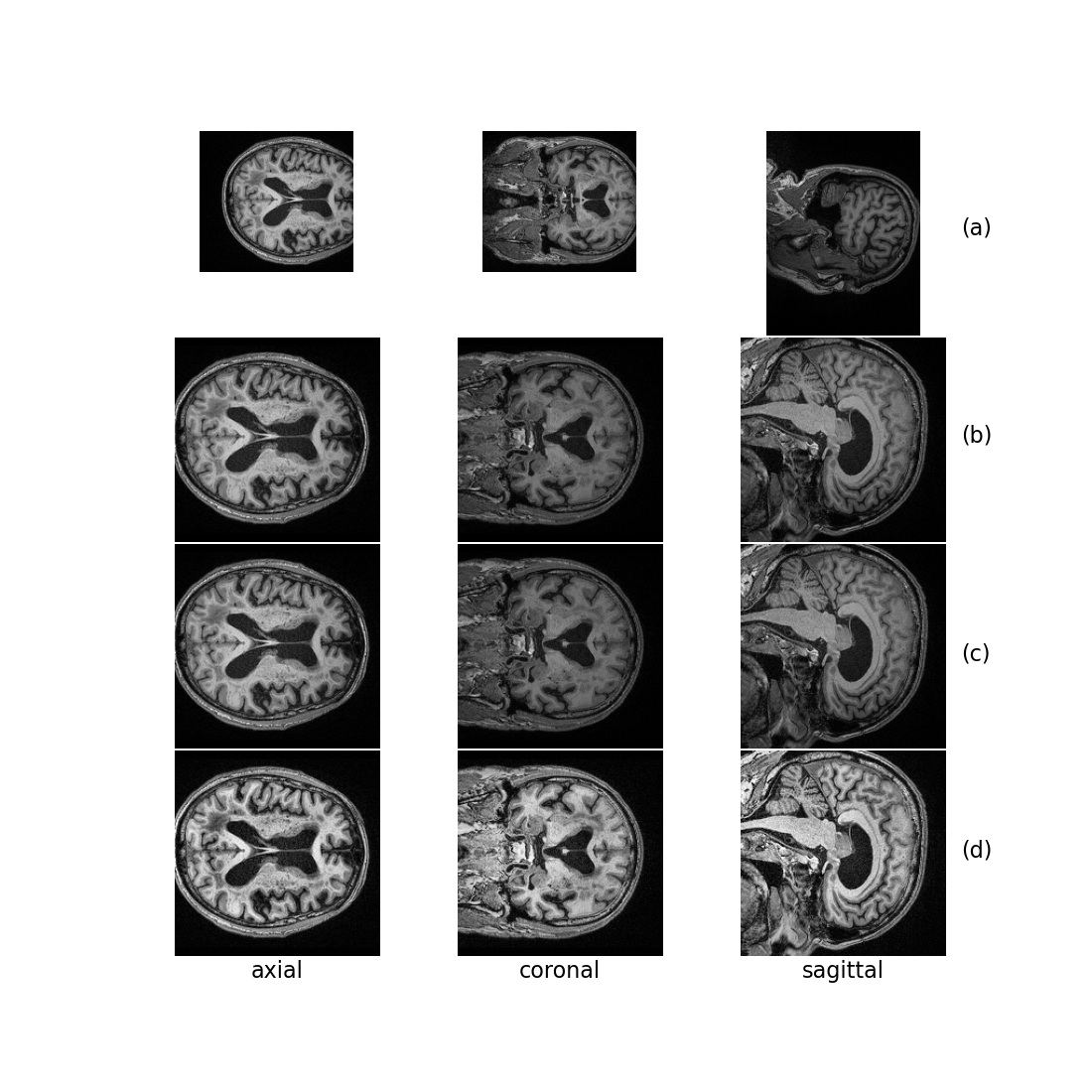}}
\vspace{-0.8cm}
\caption{Visualisation of each preprocessing step. Row (a) shows the raw image ($176\times256\times256$ voxels) of a 3D MRI scan whereas Row (b) presents the resized images ($192\times192\times192$ voxels). Row (c) shows the max-min intensity normalised image followed by row (d) of images that are processed by contrast limited adaptive histogram equalisation \cite{clahe}. }
\label{fig:preprocessing}
\end{figure}

Although many studies reported that data augmentation techniques have a significant impact on classification performance, but they often did not state a thorough explanation to support their selection of techniques. To establish a baseline of our proposed approach, this study did not apply any image augmentation techniques after preprocessing. \par

\section{Pretext Tasks}
\subsection{Brain Age Prediction}
In this work, we employed the same 3D CNN lightweight base model from our previous work \cite{zheng2022alzheimer}. The model only consists of $3$ million learnable parameters, which is more compatible with small dataset sizes and 3D volume data. \par

This model consists of 7 3D convolution blocks. Each of the first five blocks consists of a Convolutional layer, a BatchNorm layer, a MaxPool layer, and a ReLU layer. The sixth block follows the same structure but omits the MaxPooling layer. The seventh block contains an AvgPool layer, a Dropout layer, a Convolutional layer, a Flatten layer, and a Linear layer. The detail of the base model is shown in Fig.~\ref{fig:age}. The final layer is adjusted for different pretext tasks which will be illustrated later in this section. \par

\begin{figure}[htbp]
\centering
\includegraphics[width=0.975\linewidth]{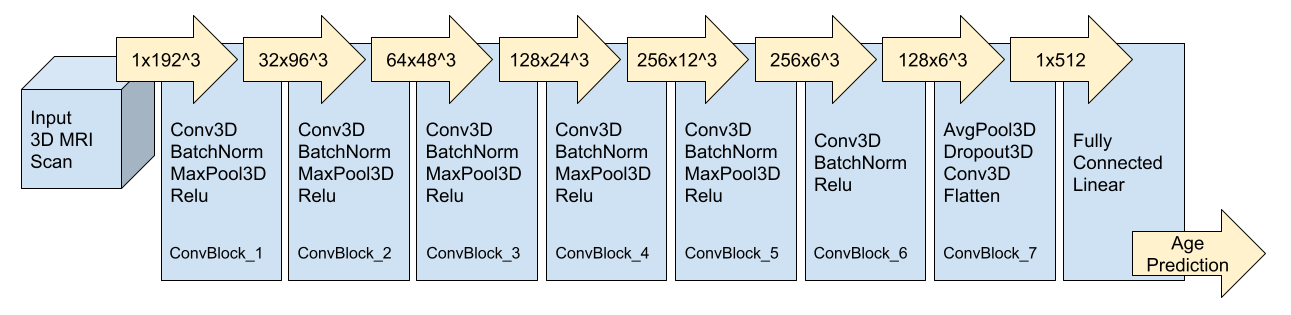}
\hfill
\caption{Overview of the proposed pretext brain age prediction task using 3D convolutional neural networks for Alzheimer's Disease detection. The base model computes the input 3D MRI scan into a numerical prediction. Note that the term $ch\times dim^{\wedge}3$ in the ``arrows" denotes the shape of intermediate data between convolutional layers. The ``$ch$" denotes the number of channels, whereas ``$dim^{\wedge}3$" indicates the size of data. All the ConvBlocks are configured to perform $3\times 3\times 3$ convolution except the $6^{th}$ block which has a $1\times 1\times 1$ kernel for downsampling purposes.} 
\label{fig:age}
\end{figure}

\subsection{Rotation Classification}
The motivation was to explore the idea of using 3D MRI rotation classification as a pretext task for feature extractor training. The idea is to engineer a pretext task by rotating an input image to certain degrees and then train the feature extractor to identify the degrees of rotation. The underlying assumption is that if an image is rotated, the semantic content of the image should remain the same, but the appearance of the image should change. \par

A 2D image can be rotated around the centre of the image for a given degree. For the human eye, the rotation can be developed in clockwise or counter-clockwise directions. Commonly the rotation of a 2D image is set to 90, 180 and 270 degrees to formulate a classification task. The 3D MRI images can be rotated in a similar way: each dimension can be seen as a stack of 2D images/slices then all dimensions are rotated around a shared centre point. Therefore, the rotation of a 3D brain image can be performed by 3 separate 2D rotations in each plane. \par

As shown in Fig.\ref{fig:rotation_example}, the first row is a 3D MRI in its original orientation, which we conveniently assign as class $0$. The number in the brackets indicates the number of 90-degree clockwise rotations performed in each plane. For example, the second row shows $(0,0,1)$ where the third dimension is rotated 90 degrees clockwise once, as class $1$. Not all 64 rotations lead to unique images, therefore only the first 32 out of 64 possible combinations of rotations are considered as classes in this work. Then the remaining 32 classes are developed accordingly. \par

\begin{figure}[htbp]
\centering
\includegraphics[width=0.975\linewidth]{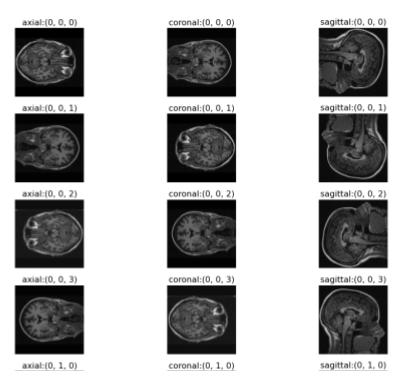}
\hfill
\caption{Examples of 3D Brain sMRI Rotation. Part (a) shows classes 1 to 8.} 
\label{fig:rotation_example}
\end{figure}

For comparison purposes, we used the same $3$ million parameter lightweight model as the feature extractor, followed by a fully connected layer and a softmax activation for classification. The details of the architecture are depicted in Fig.\ref{fig:rotation}. \par

\begin{figure}[htbp]
\centering
\includegraphics[width=0.975\linewidth]{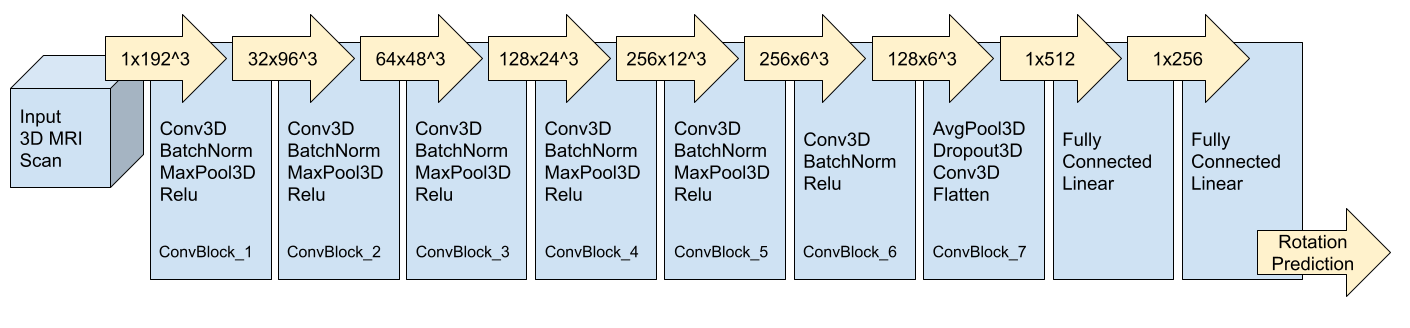}
\hfill
\caption{Overview of the proposed pretext brain rotation classification task using 3D convolutional neural networks for Alzheimer's Disease detection. The base model computes the input 3D MRI scan into a latent representation, followed by classification. Note that the term $ch\times dim^{\wedge}3$ in the ``arrows" denotes the shape of intermediate data between convolutional layers. The ``$ch$" denotes the number of channels, whereas ``$dim^{\wedge}3$" indicates the size of data. All the ConvBlocks are configured to perform $3\times 3\times 3$ convolution except the $6^{th}$ block which has a $1\times 1\times 1$ kernel for downsampling purposes.} 
\label{fig:rotation}
\end{figure}

\subsection{Image Reconstruction}
The motivation is to explore the idea of using an AutoEncoder as a self-supervised pretext task for feature extractor training. The idea is to transform the input data into a latent space representation followed by reconstruction as output. A CNN-based AutoEncoder typically utilises convolution layers in the encoder as CNN classifiers, whereas the decoder consists of deconvolution layers. The generality of the representation is learned by minimising the error between the original input and the reconstructed input. Then encoder component is utilised as a feature extractor for the downstream task. This chapter contributes towards the exploration of using AutoEncoders to learn representation for Alzheimer's Disease classification in the context of self-supervised training. \par

The encoder is identical to the CNN part of age prediction without the final numerical output layer. The flattened representations are then fed into the decoder. The detail of the proposed brain sMRI reconstruction scheme is shown in Fig.~\ref{fig:recon}. After the reconstruction training, the decoder is removed, and then the encoder is used as a feature extractor to generate representations for testing. \par

\begin{figure}[htbp]
\centering
\includegraphics[width=0.975\linewidth]{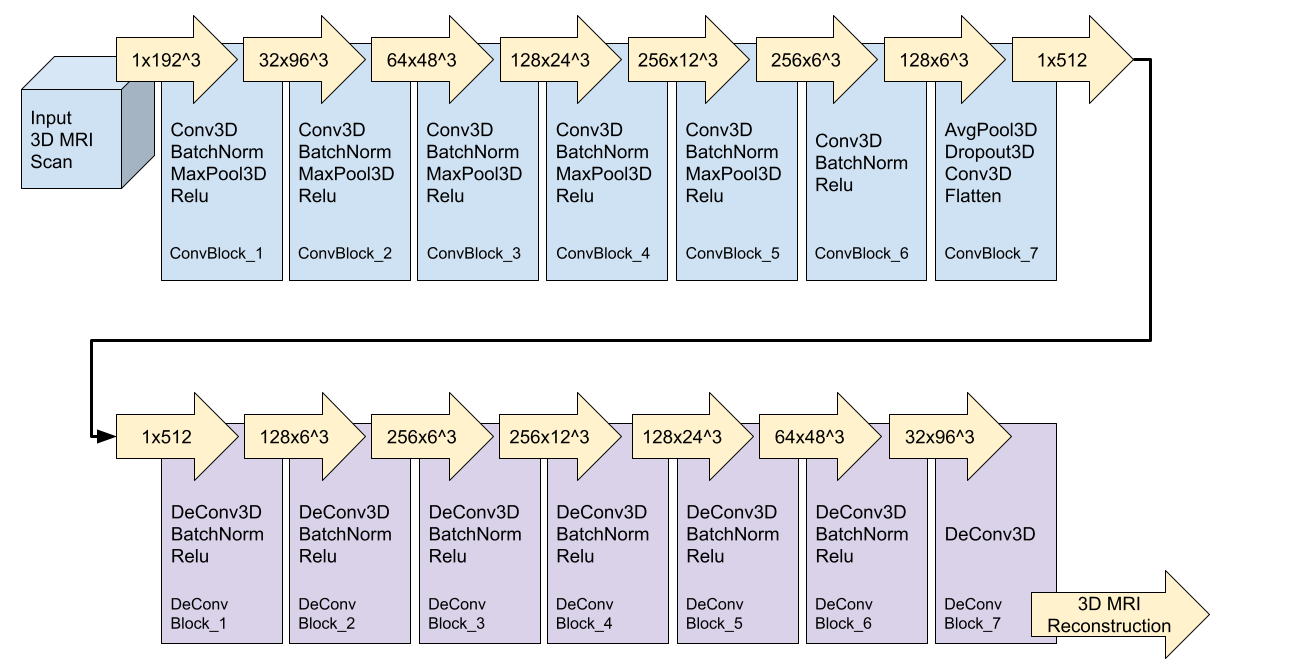}
\hfill
\caption{Overview of the proposed pretext brain image reconstruction task using 3D convolutional neural networks for Alzheimer's Disease detection. The encoder computes the input 3D MRI scan into a latent representation, followed by a decoder for reconstruction. Note that the term $ch\times dim^{\wedge}3$ in the ``arrows" denotes the shape of intermediate data between convolutional layers. The ``$ch$" denotes the number of channels, whereas ``$dim^{\wedge}3$" indicates the size of data. All the ConvBlocks are configured to perform $3\times 3\times 3$ convolution except the $6^{th}$ block which has a $1\times 1\times 1$ kernel for downsampling purposes.} 
\label{fig:recon}
\end{figure}

\subsection{Multi-Head Task}
Engineering multi-head tasks is a widely used method to utilise different modalities of inputs such as images, text, and numeric and nominal values. A multi-head task can incorporate this information for feature extractor training. However, the additional model complexity can result in significant computational costs. Therefore, the design of multi-head tasks normally takes incremental steps during the design process. This work designed the multi-head task as a combination of all the brain age prediction, rotation classification and brain image reconstruction. \par

The idea is to combine different types of output heads into one base feature extractor. In our case, we combined the brain age prediction task, the brain image reconstruction task and the brain image rotation classification task as three output heads. In the case of the reconstruction task, the output head is the decoder part. The lightweight model is employed again as the base model. While training the Multi-head task, the base feature extractor is shared between three different output heads. Each mini-batch is fed into each of the heads where the order of the head is randomly chosen during training. The parameters of the base model are updated according to each loss of the head task. The details of the architecture are depicted in Fig.\ref{fig:multihead}. \par

\begin{figure}[htbp]
\centering
\includegraphics[width=0.975\linewidth]{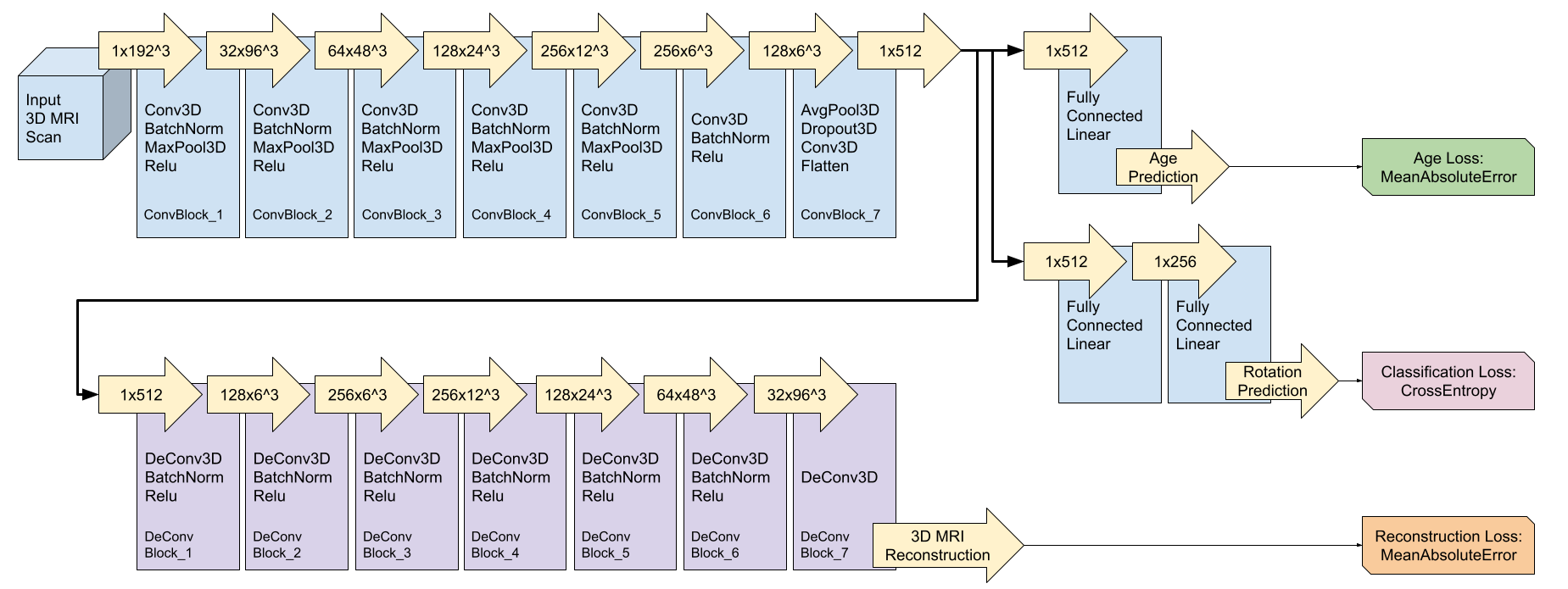}
\hfill
\caption{Overview of the proposed pretext multi-head task using 3D convolutional neural networks for Alzheimer's Disease detection. The base model compresses the input 3D MRI scan into a latent representation, followed by the brain age prediction, brain image reconstruction and brain rotation classification. Note that the term $ch\times dim^{\wedge}3$ in the ``arrows" denotes the shape of intermediate data between convolutional layers. The ``$ch$" denotes the number of channels, whereas ``$dim^{\wedge}3$" indicates the size of data. All the ConvBlocks are configured to perform $3\times 3\times 3$ convolution except the $6^{th}$ block which has a $1\times 1\times 1$ kernel for downsampling purposes.} 
\label{fig:multihead}
\end{figure}

\subsection{Representation Generation}
After the regression training, the fully connected layer is removed, and then the feature extractor portion of the 3D CNN model is used to generate representations for the scans. The usage of the feature extractor for representation generation is shown in Fig.~\ref{fig:representation}. \par

\begin{figure}[htbp]
\centering
\includegraphics[width=0.975\linewidth]{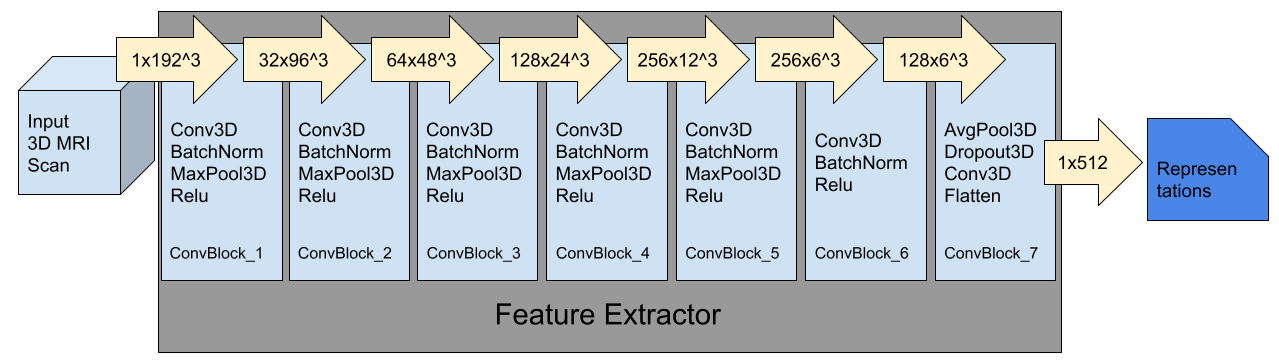}
\hfill
\caption{Overview of the proposed brain sMRI representation generation scheme using trained feature extractors.} 
\label{fig:representation}
\end{figure}

\section{Experimental Settings}
Due to the computational hardware limitation, the significantly increased cost led to the change of training strategy and hyperparameters. The model is only trained in a 1-fold setting for 50 epochs. The learning rate decays by a factor of 0.5 after every 20 epochs. The mini-batch size for each pretext task is adjusted to fit within the GPU memory constraints. Specifically, the mini-batch size for pretext age, rotation, reconstruction and multi-head are 15, 20, 12, 12, respectively. The learning rate step and gamma are configured to reduce the learning rate every 20 epochs at a rate of 0.5. The hyperparameter for training is shown in Table.~\ref{tab:hyper}. \par

Based on our preliminary experiments on various data augmentation techniques, random cropping (RandCrop) showed the best improvement in the classifications. Therefore, we employed random cropping on the fly during training. To maintain comparability to the previous chapter, the $200\times200\times200$ 3D volumes are cropped to be the shape of $192\times192\times192$. \par

\begin{table}[htbp]
\centering
\begin{tabular}{|c|c|}
\hline
\textbf{Hyperparameter} & \textbf{Value} \\ \hline
BatchSize & 15/20/12/12 \\ \hline
Epochs & 50 \\ \hline
LearningRate & 0.001 \\ \hline
LearningRate Step & 20 \\ \hline
LearningRate Gamma & 0.5 \\ \hline
\end{tabular}%
\caption{The hyperparameter setting for pretext tasks training.}
\label{tab:hyper}
\end{table}

\section{Results}
Using the LDM100K synthetic dataset for feature extractor training, the downstream AD vs. CN classifications results are shown in \Cref{tab:svc_oasis,tab:rfc_oasis,tab:svc_adni,tab:rfc_adni}. Each table presents the classification performance of 4 proposed approaches trained on the LDM100k dataset with and without random cropping using a support vector classifier or random forest classifier on either the OASIS or the ADNI dataset. \par

Among these 4 tables trained on the LDM100K dataset, the highest classification ACC of 73.9\% is achieved by the SVC classifier on features obtained by brain age prediction in conjunction with RandCrop data augmentation on the OASIS-3 dataset. The AutoEncoder with RandCrop shows a fraction less ACC but the highest SEN of 41.7\%, AUC of 0.654 and J\_stat of 0.309 while the rotation-based approach using RandCrop resulted in the highest SPE of 0.947. The RFC shows a slightly lower number of ACC (72.8\%) and AUC (0.614). A lower SEN of 31.4\% is obtained by AutoEncoder without data augmentation while a similar SPE of 96.5\% is come by the same rotation and RandCrop combination. The RFC earned a slightly poor J\_stat of 0.239 compared to SVC. Both SVC and RFC performed poorly on the ADNI testing data, which is not ideal compared with the OASIS-3 testing. One reason could be that the OASIS-3 has a higher resolution of CDR recordings that allows fine-grain subject selection. \par

\begin{table}[htbp]
\centering
\begin{tabular}{|cccccc|}
\hline
\multicolumn{6}{|c|}{\textbf{LDM Training \& OASIS Testing}} \\ \hline
\multicolumn{1}{|c|}{\textbf{SVC}} & \multicolumn{1}{c|}{\textbf{ACC}} & \multicolumn{1}{c|}{\textbf{SEN}} & \multicolumn{1}{c|}{\textbf{SPE}} & \multicolumn{1}{c|}{\textbf{AUC}} & \textbf{J\_stat} \\ \hline
\multicolumn{1}{|c|}{\textbf{Age}} & \multicolumn{1}{c|}{\cellcolor[HTML]{FFFFFF}\begin{tabular}[c]{@{}c@{}}0.707\\ +/-0.050\end{tabular}} & \multicolumn{1}{c|}{\cellcolor[HTML]{FFFFFF}\begin{tabular}[c]{@{}c@{}}0.278\\ +/-0.086\end{tabular}} & \multicolumn{1}{c|}{\cellcolor[HTML]{FFFFFF}\begin{tabular}[c]{@{}c@{}}0.912\\ +/-0.044\end{tabular}} & \multicolumn{1}{c|}{\cellcolor[HTML]{FFFFFF}\begin{tabular}[c]{@{}c@{}}0.595\\ +/-0.036\end{tabular}} & \cellcolor[HTML]{FFFFFF}\begin{tabular}[c]{@{}c@{}}0.190\\ +/-0.071\end{tabular} \\ \hline
\multicolumn{1}{|c|}{\textbf{\begin{tabular}[c]{@{}c@{}}Age\\ RandCrop\end{tabular}}} & \multicolumn{1}{c|}{\cellcolor[HTML]{FFFFFF}\textbf{\begin{tabular}[c]{@{}c@{}}0.739\\ +/-0.037\end{tabular}}} & \multicolumn{1}{c|}{\cellcolor[HTML]{FFFFFF}\begin{tabular}[c]{@{}c@{}}0.320\\ +/-0.052\end{tabular}} & \multicolumn{1}{c|}{\cellcolor[HTML]{FFFFFF}\begin{tabular}[c]{@{}c@{}}0.933\\ +/-0.025\end{tabular}} & \multicolumn{1}{c|}{\cellcolor[HTML]{FFFFFF}\begin{tabular}[c]{@{}c@{}}0.627\\ +/-0.032\end{tabular}} & \cellcolor[HTML]{FFFFFF}\begin{tabular}[c]{@{}c@{}}0.253\\ +/-0.064\end{tabular} \\ \hline
\multicolumn{1}{|c|}{\textbf{AutoEncoder}} & \multicolumn{1}{c|}{\cellcolor[HTML]{FFFFFF}\begin{tabular}[c]{@{}c@{}}0.738\\ +/-0.046\end{tabular}} & \multicolumn{1}{c|}{\cellcolor[HTML]{FFFFFF}\begin{tabular}[c]{@{}c@{}}0.402\\ +/-0.056\end{tabular}} & \multicolumn{1}{c|}{\cellcolor[HTML]{FFFFFF}\begin{tabular}[c]{@{}c@{}}0.897\\ +/-0.049\end{tabular}} & \multicolumn{1}{c|}{\cellcolor[HTML]{FFFFFF}\begin{tabular}[c]{@{}c@{}}0.649\\ +/-0.037\end{tabular}} & \cellcolor[HTML]{FFFFFF}\begin{tabular}[c]{@{}c@{}}0.299\\ +/-0.073\end{tabular} \\ \hline
\multicolumn{1}{|c|}{\textbf{\begin{tabular}[c]{@{}c@{}}AutoEncoder\\ RandCrop\end{tabular}}} & \multicolumn{1}{c|}{\cellcolor[HTML]{FFFFFF}\begin{tabular}[c]{@{}c@{}}0.738\\ +/-0.037\end{tabular}} & \multicolumn{1}{c|}{\cellcolor[HTML]{FFFFFF}\textbf{\begin{tabular}[c]{@{}c@{}}0.417\\ +/-0.088\end{tabular}}} & \multicolumn{1}{c|}{\cellcolor[HTML]{FFFFFF}\begin{tabular}[c]{@{}c@{}}0.892\\ +/-0.042\end{tabular}} & \multicolumn{1}{c|}{\cellcolor[HTML]{FFFFFF}\textbf{\begin{tabular}[c]{@{}c@{}}0.654\\ +/-0.042\end{tabular}}} & \cellcolor[HTML]{FFFFFF}\textbf{\begin{tabular}[c]{@{}c@{}}0.309\\ +/-0.084\end{tabular}} \\ \hline
\multicolumn{1}{|c|}{\textbf{Rotation}} & \multicolumn{1}{c|}{\cellcolor[HTML]{FFFFFF}\begin{tabular}[c]{@{}c@{}}0.696\\ +/-0.054\end{tabular}} & \multicolumn{1}{c|}{\cellcolor[HTML]{FFFFFF}\begin{tabular}[c]{@{}c@{}}0.187\\ +/-0.055\end{tabular}} & \multicolumn{1}{c|}{\cellcolor[HTML]{FFFFFF}\begin{tabular}[c]{@{}c@{}}0.933\\ +/-0.030\end{tabular}} & \multicolumn{1}{c|}{\cellcolor[HTML]{FFFFFF}\begin{tabular}[c]{@{}c@{}}0.560\\ +/-0.029\end{tabular}} & \cellcolor[HTML]{FFFFFF}\begin{tabular}[c]{@{}c@{}}0.120\\ +/-0.058\end{tabular} \\ \hline
\multicolumn{1}{|c|}{\textbf{\begin{tabular}[c]{@{}c@{}}Rotation\\ RandCrop\end{tabular}}} & \multicolumn{1}{c|}{\cellcolor[HTML]{FFFFFF}\begin{tabular}[c]{@{}c@{}}0.674\\ +/-0.061\end{tabular}} & \multicolumn{1}{c|}{\cellcolor[HTML]{FFFFFF}\begin{tabular}[c]{@{}c@{}}0.092\\ +/-0.068\end{tabular}} & \multicolumn{1}{c|}{\cellcolor[HTML]{FFFFFF}\textbf{\begin{tabular}[c]{@{}c@{}}0.947\\ +/-0.044\end{tabular}}} & \multicolumn{1}{c|}{\cellcolor[HTML]{FFFFFF}\begin{tabular}[c]{@{}c@{}}0.520\\ +/-0.018\end{tabular}} & \cellcolor[HTML]{FFFFFF}\begin{tabular}[c]{@{}c@{}}0.040\\ +/-0.035\end{tabular} \\ \hline
\multicolumn{1}{|c|}{\textbf{Multi-Head}} & \multicolumn{1}{c|}{\cellcolor[HTML]{FFFFFF}\begin{tabular}[c]{@{}c@{}}0.719\\ +/-0.045\end{tabular}} & \multicolumn{1}{c|}{\cellcolor[HTML]{FFFFFF}\begin{tabular}[c]{@{}c@{}}0.342\\ +/-0.082\end{tabular}} & \multicolumn{1}{c|}{\cellcolor[HTML]{FFFFFF}\begin{tabular}[c]{@{}c@{}}0.903\\ +/-0.053\end{tabular}} & \multicolumn{1}{c|}{\cellcolor[HTML]{FFFFFF}\begin{tabular}[c]{@{}c@{}}0.623\\ +/-0.041\end{tabular}} & \cellcolor[HTML]{FFFFFF}\begin{tabular}[c]{@{}c@{}}0.245\\ +/-0.083\end{tabular} \\ \hline
\multicolumn{1}{|c|}{\textbf{\begin{tabular}[c]{@{}c@{}}Multi-Head\\ RandCrop\end{tabular}}} & \multicolumn{1}{c|}{\cellcolor[HTML]{FFFFFF}\begin{tabular}[c]{@{}c@{}}0.733\\ +/-0.031\end{tabular}} & \multicolumn{1}{c|}{\cellcolor[HTML]{FFFFFF}\begin{tabular}[c]{@{}c@{}}0.379\\ +/-0.080\end{tabular}} & \multicolumn{1}{c|}{\cellcolor[HTML]{FFFFFF}\begin{tabular}[c]{@{}c@{}}0.894\\ +/-0.028\end{tabular}} & \multicolumn{1}{c|}{\cellcolor[HTML]{FFFFFF}\begin{tabular}[c]{@{}c@{}}0.636\\ +/-0.037\end{tabular}} & \cellcolor[HTML]{FFFFFF}\begin{tabular}[c]{@{}c@{}}0.272\\ +/-0.073\end{tabular} \\ \hline
\end{tabular}
\caption{The classification performance using the feature extractor trained on the LDM100K dataset via all four approaches tested on the OASIS-3 dataset using the support vector classifier.}
\label{tab:svc_oasis}
\end{table}

\begin{table}[htbp]
\begin{tabular}{|cccccc|}
\hline
\multicolumn{6}{|c|}{\textbf{LDM Training \& OASIS Testing}} \\ \hline
\multicolumn{1}{|c|}{\textbf{RFC}} & \multicolumn{1}{c|}{\textbf{ACC}} & \multicolumn{1}{c|}{\textbf{SEN}} & \multicolumn{1}{c|}{\textbf{SPE}} & \multicolumn{1}{c|}{\textbf{AUC}} & \textbf{J\_stat} \\ \hline
\multicolumn{1}{|c|}{\textbf{Age}} & \multicolumn{1}{c|}{\cellcolor[HTML]{FFFFFF}\begin{tabular}[c]{@{}c@{}}0.724\\ +/-0.044\end{tabular}} & \multicolumn{1}{c|}{\cellcolor[HTML]{FFFFFF}\begin{tabular}[c]{@{}c@{}}0.321\\ +/-0.056\end{tabular}} & \multicolumn{1}{c|}{\cellcolor[HTML]{FFFFFF}\begin{tabular}[c]{@{}c@{}}0.919\\ +/-0.044\end{tabular}} & \multicolumn{1}{c|}{\cellcolor[HTML]{FFFFFF}\begin{tabular}[c]{@{}c@{}}0.620\\ +/-0.026\end{tabular}} & \cellcolor[HTML]{FFFFFF}\textbf{\begin{tabular}[c]{@{}c@{}}0.239\\ +/-0.051\end{tabular}} \\ \hline
\multicolumn{1}{|c|}{\textbf{\begin{tabular}[c]{@{}c@{}}Age\\ Randdcrop\end{tabular}}} & \multicolumn{1}{c|}{\cellcolor[HTML]{FFFFFF}\textbf{\begin{tabular}[c]{@{}c@{}}0.728\\ +/-0.052\end{tabular}}} & \multicolumn{1}{c|}{\cellcolor[HTML]{FFFFFF}\begin{tabular}[c]{@{}c@{}}0.304\\ +/-0.066\end{tabular}} & \multicolumn{1}{c|}{\cellcolor[HTML]{FFFFFF}\begin{tabular}[c]{@{}c@{}}0.925\\ +/-0.021\end{tabular}} & \multicolumn{1}{c|}{\cellcolor[HTML]{FFFFFF}\textbf{\begin{tabular}[c]{@{}c@{}}0.614\\ +/-0.035\end{tabular}}} & \cellcolor[HTML]{FFFFFF}\begin{tabular}[c]{@{}c@{}}0.228\\ +/-0.071\end{tabular} \\ \hline
\multicolumn{1}{|c|}{\textbf{AutoEncoder}} & \multicolumn{1}{c|}{\cellcolor[HTML]{FFFFFF}\begin{tabular}[c]{@{}c@{}}0.721\\ +/-0.045\end{tabular}} & \multicolumn{1}{c|}{\cellcolor[HTML]{FFFFFF}\textbf{\begin{tabular}[c]{@{}c@{}}0.314\\ +/-0.093\end{tabular}}} & \multicolumn{1}{c|}{\cellcolor[HTML]{FFFFFF}\begin{tabular}[c]{@{}c@{}}0.909\\ +/-0.051\end{tabular}} & \multicolumn{1}{c|}{\cellcolor[HTML]{FFFFFF}\begin{tabular}[c]{@{}c@{}}0.612\\ +/-0.040\end{tabular}} & \cellcolor[HTML]{FFFFFF}\begin{tabular}[c]{@{}c@{}}0.223\\ +/-0.079\end{tabular} \\ \hline
\multicolumn{1}{|c|}{\textbf{\begin{tabular}[c]{@{}c@{}}AutoEncoder\\ Randdcrop\end{tabular}}} & \multicolumn{1}{c|}{\cellcolor[HTML]{FFFFFF}\begin{tabular}[c]{@{}c@{}}0.712\\ +/-0.050\end{tabular}} & \multicolumn{1}{c|}{\cellcolor[HTML]{FFFFFF}\begin{tabular}[c]{@{}c@{}}0.266\\ +/-0.096\end{tabular}} & \multicolumn{1}{c|}{\cellcolor[HTML]{FFFFFF}\begin{tabular}[c]{@{}c@{}}0.927\\ +/-0.052\end{tabular}} & \multicolumn{1}{c|}{\cellcolor[HTML]{FFFFFF}\begin{tabular}[c]{@{}c@{}}0.597\\ +/-0.037\end{tabular}} & \cellcolor[HTML]{FFFFFF}\begin{tabular}[c]{@{}c@{}}0.194\\ +/-0.075\end{tabular} \\ \hline
\multicolumn{1}{|c|}{\textbf{Rotation}} & \multicolumn{1}{c|}{\cellcolor[HTML]{FFFFFF}\begin{tabular}[c]{@{}c@{}}0.701\\ +/-0.055\end{tabular}} & \multicolumn{1}{c|}{\cellcolor[HTML]{FFFFFF}\begin{tabular}[c]{@{}c@{}}0.163\\ +/-0.047\end{tabular}} & \multicolumn{1}{c|}{\cellcolor[HTML]{FFFFFF}\begin{tabular}[c]{@{}c@{}}0.954\\ +/-0.027\end{tabular}} & \multicolumn{1}{c|}{\cellcolor[HTML]{FFFFFF}\begin{tabular}[c]{@{}c@{}}0.559\\ +/-0.026\end{tabular}} & \cellcolor[HTML]{FFFFFF}\begin{tabular}[c]{@{}c@{}}0.117\\ +/-0.051\end{tabular} \\ \hline
\multicolumn{1}{|c|}{\textbf{\begin{tabular}[c]{@{}c@{}}Rotation\\ Randdcrop\end{tabular}}} & \multicolumn{1}{c|}{\cellcolor[HTML]{FFFFFF}\begin{tabular}[c]{@{}c@{}}0.692\\ +/-0.058\end{tabular}} & \multicolumn{1}{c|}{\cellcolor[HTML]{FFFFFF}\begin{tabular}[c]{@{}c@{}}0.102\\ +/-0.041\end{tabular}} & \multicolumn{1}{c|}{\cellcolor[HTML]{FFFFFF}\textbf{\begin{tabular}[c]{@{}c@{}}0.965\\ +/-0.024\end{tabular}}} & \multicolumn{1}{c|}{\cellcolor[HTML]{FFFFFF}\begin{tabular}[c]{@{}c@{}}0.534\\ +/-0.013\end{tabular}} & \cellcolor[HTML]{FFFFFF}\begin{tabular}[c]{@{}c@{}}0.067\\ +/-0.026\end{tabular} \\ \hline
\multicolumn{1}{|c|}{\textbf{Multi-Head}} & \multicolumn{1}{c|}{\cellcolor[HTML]{FFFFFF}\begin{tabular}[c]{@{}c@{}}0.684\\ +/-0.052\end{tabular}} & \multicolumn{1}{c|}{\cellcolor[HTML]{FFFFFF}\begin{tabular}[c]{@{}c@{}}0.133\\ +/-0.066\end{tabular}} & \multicolumn{1}{c|}{\cellcolor[HTML]{FFFFFF}\begin{tabular}[c]{@{}c@{}}0.944\\ +/-0.039\end{tabular}} & \multicolumn{1}{c|}{\cellcolor[HTML]{FFFFFF}\begin{tabular}[c]{@{}c@{}}0.539\\ +/-0.021\end{tabular}} & \cellcolor[HTML]{FFFFFF}\begin{tabular}[c]{@{}c@{}}0.077\\ +/-0.041\end{tabular} \\ \hline
\multicolumn{1}{|c|}{\textbf{\begin{tabular}[c]{@{}c@{}}Multi-Head\\ RandCrop\end{tabular}}} & \multicolumn{1}{c|}{\cellcolor[HTML]{FFFFFF}\begin{tabular}[c]{@{}c@{}}0.702\\ +/-0.053\end{tabular}} & \multicolumn{1}{c|}{\cellcolor[HTML]{FFFFFF}\begin{tabular}[c]{@{}c@{}}0.243\\ +/-0.067\end{tabular}} & \multicolumn{1}{c|}{\cellcolor[HTML]{FFFFFF}\begin{tabular}[c]{@{}c@{}}0.910\\ +/-0.039\end{tabular}} & \multicolumn{1}{c|}{\cellcolor[HTML]{FFFFFF}\begin{tabular}[c]{@{}c@{}}0.577\\ +/-0.025\end{tabular}} & \cellcolor[HTML]{FFFFFF}\begin{tabular}[c]{@{}c@{}}0.154\\ +/-0.049\end{tabular} \\ \hline
\end{tabular}
\caption{The classification performance using the feature extractor trained on the LDM100K dataset via all four approaches tested on the OASIS-3 dataset using the random forest classifier.}
\label{tab:rfc_oasis}
\end{table}

\begin{table}[htbp]
\begin{tabular}{|cccccc|}
\hline
\multicolumn{6}{|c|}{\textbf{LDM Training \& ADNI Testing}} \\ \hline
\multicolumn{1}{|c|}{\textbf{SVC}} & \multicolumn{1}{c|}{\textbf{ACC}} & \multicolumn{1}{c|}{\textbf{SEN}} & \multicolumn{1}{c|}{\textbf{SPE}} & \multicolumn{1}{c|}{\textbf{AUC}} & \textbf{J\_stat} \\ \hline
\multicolumn{1}{|c|}{\textbf{Age}} & \multicolumn{1}{c|}{\cellcolor[HTML]{FFFFFF}\begin{tabular}[c]{@{}c@{}}0.653\\ +/-0.063\end{tabular}} & \multicolumn{1}{c|}{\cellcolor[HTML]{FFFFFF}\begin{tabular}[c]{@{}c@{}}0.000\\ +/-0.000\end{tabular}} & \multicolumn{1}{c|}{\cellcolor[HTML]{FFFFFF}\textbf{\begin{tabular}[c]{@{}c@{}}1.000\\ +/-0.000\end{tabular}}} & \multicolumn{1}{c|}{\cellcolor[HTML]{FFFFFF}\begin{tabular}[c]{@{}c@{}}0.500\\ +/-0.000\end{tabular}} & \cellcolor[HTML]{FFFFFF}\begin{tabular}[c]{@{}c@{}}0.000\\ +/-0.000\end{tabular} \\ \hline
\multicolumn{1}{|c|}{\textbf{\begin{tabular}[c]{@{}c@{}}Age\\ RandCrop\end{tabular}}} & \multicolumn{1}{c|}{\cellcolor[HTML]{FFFFFF}\begin{tabular}[c]{@{}c@{}}0.643\\ +/-0.091\end{tabular}} & \multicolumn{1}{c|}{\cellcolor[HTML]{FFFFFF}\textbf{\begin{tabular}[c]{@{}c@{}}0.175\\ +/-0.285\end{tabular}}} & \multicolumn{1}{c|}{\cellcolor[HTML]{FFFFFF}\begin{tabular}[c]{@{}c@{}}0.865\\ +/-0.290\end{tabular}} & \multicolumn{1}{c|}{\cellcolor[HTML]{FFFFFF}\begin{tabular}[c]{@{}c@{}}0.520\\ +/-0.033\end{tabular}} & \cellcolor[HTML]{FFFFFF}\begin{tabular}[c]{@{}c@{}}0.040\\ +/-0.067\end{tabular} \\ \hline
\multicolumn{1}{|c|}{\textbf{Rotation}} & \multicolumn{1}{c|}{\cellcolor[HTML]{FFFFFF}\begin{tabular}[c]{@{}c@{}}0.649\\ +/-0.062\end{tabular}} & \multicolumn{1}{c|}{\cellcolor[HTML]{FFFFFF}\begin{tabular}[c]{@{}c@{}}0.000\\ +/-0.000\end{tabular}} & \multicolumn{1}{c|}{\cellcolor[HTML]{FFFFFF}\begin{tabular}[c]{@{}c@{}}0.994\\ +/-0.017\end{tabular}} & \multicolumn{1}{c|}{\cellcolor[HTML]{FFFFFF}\begin{tabular}[c]{@{}c@{}}0.497\\ +/-0.008\end{tabular}} & \cellcolor[HTML]{FFFFFF}\begin{tabular}[c]{@{}c@{}}-0.006\\ +/-0.017\end{tabular} \\ \hline
\multicolumn{1}{|c|}{\textbf{\begin{tabular}[c]{@{}c@{}}Rotation\\ RandCrop\end{tabular}}} & \multicolumn{1}{c|}{\cellcolor[HTML]{FFFFFF}\begin{tabular}[c]{@{}c@{}}0.651\\ +/-0.064\end{tabular}} & \multicolumn{1}{c|}{\cellcolor[HTML]{FFFFFF}\begin{tabular}[c]{@{}c@{}}0.007\\ +/-0.015\end{tabular}} & \multicolumn{1}{c|}{\cellcolor[HTML]{FFFFFF}\begin{tabular}[c]{@{}c@{}}0.993\\ +/-0.015\end{tabular}} & \multicolumn{1}{c|}{\cellcolor[HTML]{FFFFFF}\begin{tabular}[c]{@{}c@{}}0.500\\ +/-0.005\end{tabular}} & \cellcolor[HTML]{FFFFFF}\begin{tabular}[c]{@{}c@{}}0.001\\ +/-0.010\end{tabular} \\ \hline
\multicolumn{1}{|c|}{\textbf{AutoEncoder}} & \multicolumn{1}{c|}{\cellcolor[HTML]{FFFFFF}\begin{tabular}[c]{@{}c@{}}0.653\\ +/-0.063\end{tabular}} & \multicolumn{1}{c|}{\cellcolor[HTML]{FFFFFF}\begin{tabular}[c]{@{}c@{}}0.000\\ +/-0.000\end{tabular}} & \multicolumn{1}{c|}{\cellcolor[HTML]{FFFFFF}\textbf{\begin{tabular}[c]{@{}c@{}}1.000\\ +/-0.000\end{tabular}}} & \multicolumn{1}{c|}{\cellcolor[HTML]{FFFFFF}\begin{tabular}[c]{@{}c@{}}0.500\\ +/-0.000\end{tabular}} & \cellcolor[HTML]{FFFFFF}\begin{tabular}[c]{@{}c@{}}0.000\\ +/-0.000\end{tabular} \\ \hline
\multicolumn{1}{|c|}{\textbf{\begin{tabular}[c]{@{}c@{}}AutoEncoder\\ RandCrop\end{tabular}}} & \multicolumn{1}{c|}{\cellcolor[HTML]{FFFFFF}\begin{tabular}[c]{@{}c@{}}0.653\\ +/-0.063\end{tabular}} & \multicolumn{1}{c|}{\cellcolor[HTML]{FFFFFF}\begin{tabular}[c]{@{}c@{}}0.000\\ +/-0.000\end{tabular}} & \multicolumn{1}{c|}{\cellcolor[HTML]{FFFFFF}\textbf{\begin{tabular}[c]{@{}c@{}}1.000\\ +/-0.000\end{tabular}}} & \multicolumn{1}{c|}{\cellcolor[HTML]{FFFFFF}\begin{tabular}[c]{@{}c@{}}0.500\\ +/-0.000\end{tabular}} & \cellcolor[HTML]{FFFFFF}\begin{tabular}[c]{@{}c@{}}0.000\\ +/-0.000\end{tabular} \\ \hline
\multicolumn{1}{|c|}{\textbf{Multi-Head}} & \multicolumn{1}{c|}{\cellcolor[HTML]{FFFFFF}\begin{tabular}[c]{@{}c@{}}0.618\\ +/-0.068\end{tabular}} & \multicolumn{1}{c|}{\cellcolor[HTML]{FFFFFF}\begin{tabular}[c]{@{}c@{}}0.057\\ +/-0.117\end{tabular}} & \multicolumn{1}{c|}{\cellcolor[HTML]{FFFFFF}\begin{tabular}[c]{@{}c@{}}0.927\\ +/-0.131\end{tabular}} & \multicolumn{1}{c|}{\cellcolor[HTML]{FFFFFF}\begin{tabular}[c]{@{}c@{}}0.492\\ +/-0.040\end{tabular}} & \cellcolor[HTML]{FFFFFF}\begin{tabular}[c]{@{}c@{}}-0.016\\ +/-0.079\end{tabular} \\ \hline
\multicolumn{1}{|c|}{\textbf{\begin{tabular}[c]{@{}c@{}}Multi-Head\\ RandCrop\end{tabular}}} & \multicolumn{1}{c|}{\cellcolor[HTML]{FFFFFF}\textbf{\begin{tabular}[c]{@{}c@{}}0.661\\ +/-0.052\end{tabular}}} & \multicolumn{1}{c|}{\cellcolor[HTML]{FFFFFF}\begin{tabular}[c]{@{}c@{}}0.144\\ +/-0.104\end{tabular}} & \multicolumn{1}{c|}{\cellcolor[HTML]{FFFFFF}\begin{tabular}[c]{@{}c@{}}0.930\\ +/-0.047\end{tabular}} & \multicolumn{1}{c|}{\cellcolor[HTML]{FFFFFF}\textbf{\begin{tabular}[c]{@{}c@{}}0.537\\ +/-0.038\end{tabular}}} & \cellcolor[HTML]{FFFFFF}\textbf{\begin{tabular}[c]{@{}c@{}}0.075\\ +/-0.075\end{tabular}} \\ \hline
\end{tabular}
\caption{The classification performance using the feature extractor trained on the LDM100K dataset via all four approaches tested on the ADNI dataset using the support vector classifier.}
\label{tab:svc_adni}
\end{table}

\begin{table}[htbp]
\begin{tabular}{|cccccc|}
\hline
\multicolumn{6}{|c|}{\textbf{LDM Training \& ADNI Testing}} \\ \hline
\multicolumn{1}{|c|}{\textbf{RFC}} & \multicolumn{1}{c|}{\cellcolor[HTML]{FFFFFF}\textbf{ACC}} & \multicolumn{1}{c|}{\cellcolor[HTML]{FFFFFF}\textbf{SEN}} & \multicolumn{1}{c|}{\cellcolor[HTML]{FFFFFF}\textbf{SPE}} & \multicolumn{1}{c|}{\cellcolor[HTML]{FFFFFF}\textbf{AUC}} & \cellcolor[HTML]{FFFFFF}\textbf{J\_stat} \\ \hline
\multicolumn{1}{|c|}{\textbf{Age}} & \multicolumn{1}{c|}{\cellcolor[HTML]{FFFFFF}\textbf{\begin{tabular}[c]{@{}c@{}}0.653\\ +/-0.063\end{tabular}}} & \multicolumn{1}{c|}{\cellcolor[HTML]{FFFFFF}\begin{tabular}[c]{@{}c@{}}0.000\\ +/-0.000\end{tabular}} & \multicolumn{1}{c|}{\cellcolor[HTML]{FFFFFF}\textbf{\begin{tabular}[c]{@{}c@{}}1.000\\ +/-0.000\end{tabular}}} & \multicolumn{1}{c|}{\cellcolor[HTML]{FFFFFF}\begin{tabular}[c]{@{}c@{}}0.500\\ +/-0.000\end{tabular}} & \cellcolor[HTML]{FFFFFF}\begin{tabular}[c]{@{}c@{}}0.000\\ +/-0.000\end{tabular} \\ \hline
\multicolumn{1}{|c|}{\textbf{\begin{tabular}[c]{@{}c@{}}Age\\ RandCrop\end{tabular}}} & \multicolumn{1}{c|}{\cellcolor[HTML]{FFFFFF}\begin{tabular}[c]{@{}c@{}}0.635\\ +/-0.077\end{tabular}} & \multicolumn{1}{c|}{\cellcolor[HTML]{FFFFFF}\begin{tabular}[c]{@{}c@{}}0.009\\ +/-0.020\end{tabular}} & \multicolumn{1}{c|}{\cellcolor[HTML]{FFFFFF}\begin{tabular}[c]{@{}c@{}}0.966\\ +/-0.037\end{tabular}} & \multicolumn{1}{c|}{\cellcolor[HTML]{FFFFFF}\begin{tabular}[c]{@{}c@{}}0.488\\ +/-0.021\end{tabular}} & \cellcolor[HTML]{FFFFFF}\begin{tabular}[c]{@{}c@{}}-0.025\\ +/-0.041\end{tabular} \\ \hline
\multicolumn{1}{|c|}{\textbf{Rotation}} & \multicolumn{1}{c|}{\cellcolor[HTML]{FFFFFF}\begin{tabular}[c]{@{}c@{}}0.643\\ +/-0.059\end{tabular}} & \multicolumn{1}{c|}{\cellcolor[HTML]{FFFFFF}\begin{tabular}[c]{@{}c@{}}0.003\\ +/-0.010\end{tabular}} & \multicolumn{1}{c|}{\cellcolor[HTML]{FFFFFF}\begin{tabular}[c]{@{}c@{}}0.983\\ +/-0.023\end{tabular}} & \multicolumn{1}{c|}{\cellcolor[HTML]{FFFFFF}\begin{tabular}[c]{@{}c@{}}0.493\\ +/-0.014\end{tabular}} & \cellcolor[HTML]{FFFFFF}\begin{tabular}[c]{@{}c@{}}-0.014\\ +/-0.028\end{tabular} \\ \hline
\multicolumn{1}{|c|}{\textbf{\begin{tabular}[c]{@{}c@{}}Rotation\\ RandCrop\end{tabular}}} & \multicolumn{1}{c|}{\cellcolor[HTML]{FFFFFF}\begin{tabular}[c]{@{}c@{}}0.651\\ +/-0.060\end{tabular}} & \multicolumn{1}{c|}{\cellcolor[HTML]{FFFFFF}\begin{tabular}[c]{@{}c@{}}0.013\\ +/-0.021\end{tabular}} & \multicolumn{1}{c|}{\cellcolor[HTML]{FFFFFF}\begin{tabular}[c]{@{}c@{}}0.990\\ +/-0.019\end{tabular}} & \multicolumn{1}{c|}{\cellcolor[HTML]{FFFFFF}\textbf{\begin{tabular}[c]{@{}c@{}}0.502\\ +/-0.014\end{tabular}}} & \cellcolor[HTML]{FFFFFF}\textbf{\begin{tabular}[c]{@{}c@{}}0.004\\ +/-0.029\end{tabular}} \\ \hline
\multicolumn{1}{|c|}{\textbf{AutoEncoder}} & \multicolumn{1}{c|}{\cellcolor[HTML]{FFFFFF}\textbf{\begin{tabular}[c]{@{}c@{}}0.653\\ +/-0.063\end{tabular}}} & \multicolumn{1}{c|}{\cellcolor[HTML]{FFFFFF}\begin{tabular}[c]{@{}c@{}}0.000\\ +/-0.000\end{tabular}} & \multicolumn{1}{c|}{\cellcolor[HTML]{FFFFFF}\textbf{\begin{tabular}[c]{@{}c@{}}1.000\\ +/-0.000\end{tabular}}} & \multicolumn{1}{c|}{\cellcolor[HTML]{FFFFFF}\begin{tabular}[c]{@{}c@{}}0.500\\ +/-0.000\end{tabular}} & \cellcolor[HTML]{FFFFFF}\begin{tabular}[c]{@{}c@{}}0.000\\ +/-0.000\end{tabular} \\ \hline
\multicolumn{1}{|c|}{\textbf{\begin{tabular}[c]{@{}c@{}}AutoEncoder\\ RandCrop\end{tabular}}} & \multicolumn{1}{c|}{\cellcolor[HTML]{FFFFFF}\begin{tabular}[c]{@{}c@{}}0.647\\ +/-0.073\end{tabular}} & \multicolumn{1}{c|}{\cellcolor[HTML]{FFFFFF}\begin{tabular}[c]{@{}c@{}}0.000\\ +/-0.000\end{tabular}} & \multicolumn{1}{c|}{\cellcolor[HTML]{FFFFFF}\begin{tabular}[c]{@{}c@{}}0.989\\ +/-0.032\end{tabular}} & \multicolumn{1}{c|}{\cellcolor[HTML]{FFFFFF}\begin{tabular}[c]{@{}c@{}}0.495\\ +/-0.016\end{tabular}} & \cellcolor[HTML]{FFFFFF}\begin{tabular}[c]{@{}c@{}}-0.011\\ +/-0.032\end{tabular} \\ \hline
\multicolumn{1}{|c|}{\textbf{Multi-Head}} & \multicolumn{1}{c|}{\cellcolor[HTML]{FFFFFF}\begin{tabular}[c]{@{}c@{}}0.616\\ +/-0.051\end{tabular}} & \multicolumn{1}{c|}{\cellcolor[HTML]{FFFFFF}\textbf{\begin{tabular}[c]{@{}c@{}}0.021\\ +/-0.027\end{tabular}}} & \multicolumn{1}{c|}{\cellcolor[HTML]{FFFFFF}\begin{tabular}[c]{@{}c@{}}0.938\\ +/-0.080\end{tabular}} & \multicolumn{1}{c|}{\cellcolor[HTML]{FFFFFF}\begin{tabular}[c]{@{}c@{}}0.479\\ +/-0.035\end{tabular}} & \cellcolor[HTML]{FFFFFF}\begin{tabular}[c]{@{}c@{}}-0.041\\ +/-0.070\end{tabular} \\ \hline
\multicolumn{1}{|c|}{\textbf{\begin{tabular}[c]{@{}c@{}}Multi-Head\\ RandCrop\end{tabular}}} & \multicolumn{1}{c|}{\cellcolor[HTML]{FFFFFF}\begin{tabular}[c]{@{}c@{}}0.638\\ +/-0.063\end{tabular}} & \multicolumn{1}{c|}{\cellcolor[HTML]{FFFFFF}\begin{tabular}[c]{@{}c@{}}0.016\\ +/-0.017\end{tabular}} & \multicolumn{1}{c|}{\cellcolor[HTML]{FFFFFF}\begin{tabular}[c]{@{}c@{}}0.968\\ +/-0.032\end{tabular}} & \multicolumn{1}{c|}{\cellcolor[HTML]{FFFFFF}\begin{tabular}[c]{@{}c@{}}0.492\\ +/-0.019\end{tabular}} & \cellcolor[HTML]{FFFFFF}\begin{tabular}[c]{@{}c@{}}-0.016\\ +/-0.039\end{tabular} \\ \hline
\end{tabular}
\caption{The classification performance using the feature extractor trained on the LDM100K dataset via all four approaches tested on the ADNI dataset using the random forest classifier.}
\label{tab:rfc_adni}
\end{table}

\section{Discussion and Conclusion}
As shown in \cref{tab:svc_oasis,tab:svc_adni,tab:rfc_oasis,tab:rfc_adni}, 
the brain age prediction pretext task shows the best AD classification accuracy across the board. It is not as excellent as some approaches in the literature, but it is has a lot of potential to be explored in the future in terms of advanced regression models and training techniques. \par

The results of utilising the synthetic LDM-100k dataset are not as good as using real-world data, especially since the sensitivity is low. This might be due to the extreme class imbalance in the dataset. The CN class is the majority up to 90\% in the train set. Even though the test set is close to the 1:1 class ratio, the model is biased toward the majority class during prediction. \par

Future work is needed to address this severe imbalance in the train set. For example, upsampling of the minority while downsampling the majority class is a widely used technique during model training, it might help mitigate the impact of imbalance and improve sensitivity. Adjusting class weights is another popular method to address the class imbalance issue. Also, data augmentation could be another choice to leverage the minority class. Last but not least, synthetic data generation for the AD class has not yet been explored either. \par

\section{Limitations \& Future Work}
Although the evaluation results support the feasibility of the proposed approaches, there are some limitations. Firstly, the architecture of the 3D CNN base model, such as the number and types of convolutional layers, may not be the optimal choice. The application of more complex models is limited by the availability of computational hardware. Secondly, the proposed pretext tasks are only using T1w structural MRI data for feature extractor training, which leaves other available neuroimaging modalities underutilised. Last but not least, The temporal relationship between the MRI scans is underutilised. \par

This study obtained experimental results that open possibilities for future research. The potential of using more complex base models such ResidualNetwork or VisionTransformer can be further investigated. There are data available in other neuroimaging modalities (e.g. T2w, fMRI, PET, etc) and demographic information (e.g. gender, education, occupation, etc) that can be further explored. The similarities between two adjacent MRI scans might be used to enforce an AutoEncoder to minimise as part of the loss function. It is also intriguing to further explore the temporal relationship, which might lead to a more comprehensive representation of AD development and progression. \par

Currently, many computational software are developed for 2D inputs. Dedicated software libraries for 3D neuroimaging data are still in its infancy. Also, specialised hardware for 3D neuroimaging data is either extremely expensive to buy, very difficult to access or even does not exist. Therefore, advancing computational software and hardware for 3D neuroimaging data would be another interesting direction for future research. \par

In conclusion, all four pretext tasks in this study are based on the idea that using self-supervised learning methods can improve the feature extractor for Alzheimer's Disease classification performance. This study solely scratched the surface of the self-supervised learning field regarding AD classification. However, self-supervised learning might be the most promising choice of approach when neuroimaging data labelling or segmentation is unavailable. \par

\bibliographystyle{IEEEtran}
\bibliography{IEEEabrv,refs_old,refs_new,refs_ssl}

\end{document}